\documentclass[aps,amsmath,amssymb,twocolumn,showpacs,longbibliography]{revtex4-1}

\usepackage{graphicx}
\usepackage{bm}
\usepackage{color}

\begin{document}
\title{Transmission eigenchannels for coherent phonon transport}
\author{J. C. Kl\"{o}ckner$^{1,3}$}
\email{Jan.Kloeckner@uni-konstanz.de}
\author{J. C. Cuevas$^{1,2}$}
\author{F. Pauly$^{1,3}$}

\affiliation{$^{1}$Department of Physics, University of Konstanz, D-78457
  Konstanz, Germany}
\affiliation{$^{2}$Departamento de F\'{\i}sica Te\'orica
  de la Materia Condensada and Condensed Matter Physics Center (IFIMAC),
  Universidad Aut\'onoma de Madrid, E-28049 Madrid, Spain}
\affiliation{$^{3}$Okinawa Institute of Science and Technology Graduate
  University, Onna-son, Okinawa 904-0395, Japan}
  
\date{\today}

\begin{abstract}
We present a procedure to determine transmission eigenchannels for coherent
phonon transport in nanoscale devices using the framework of nonequilibrium
Green's functions. We illustrate our procedure by analyzing a one-dimensional
chain, where all steps can be carried out analytically. More importantly, we
show how the procedure can be combined with ab initio calculations to provide
a better understanding of phonon heat transport in realistic atomic-scale
junctions. In particular, we study the phonon eigenchannels in a gold metallic
atomic-size contact and different single-molecule junctions based on molecules
such as an alkane chain, C$_{60}$, and a brominated benzene-diamine, where in
this latter case destructive phonon interference effects take place.
\end{abstract}

\maketitle

\section{Introduction}

Recent advances in experimental techniques have enabled to explore the heat
conduction in a great variety of nanoscale systems
\cite{Cahill2003,Pop2010,Luo2013,Cahill2014}. It has even become possible to
measure the heat conductance of metallic wires all the way down to single-atom
contacts \cite{Cui2017,Mosso2017}, which constitute the ultimate limit of
miniaturization of electronic and phononic systems. Research on heat
conduction in nanoscale devices allows us to investigate the phonon transport
in new regimes, where the theoretical description often requires fully
atomistic approaches \cite{Minnich2015}.  Here we are especially interested in
the theoretical analysis of phonon transport in atomic and molecular
junctions, which are prototypical nanosystems that are studied intensely in
the field of molecular electronics \cite{Cuevas2017}.  In these atomic-scale
systems, the inelastic mean free path for phonons is often much larger than
the junction dimensions, and the phonon transport is therefore fully
coherent. In this situation, the phonon transport is described within the
framework of the Landauer-B\"uttiker scattering theory in which the
contribution to the thermal conductance is determined by the elastic phonon
transmission function of the system \cite{Cuevas2017,Segal2016}. Different
strategies have been put forward to compute this transmission function based
on, for instance, the scattering matrix approach
\cite{Fagas1999,Zhao2005,Wang2006,Duchemin2011,Zhang2011}, mode matching
theory \cite{Tanaka2005,Antonyuk2005,Murphy2007}, or nonequilibrium Green's
function (NEGF) techniques
\cite{Mingo2003,Mingo2006,Asai2008,Markussen2013,Sadeghi2015,Buerkle2015,Li2015,Li2017,Buerkle2017,Famili2017}.
These approaches are nicely summarized in Ref.~\cite{Wang2008}.

In the context of electronic transport, it has been shown that one can obtain
a deep insight by resolving the total transmission $\tau=\sum_{\mu}\tau_\mu$
into contributions of eigenchannels, which are particular scattering states
with transmission coefficients $0 \le \tau_\mu \le 1$. The analysis of the
eigenchannels in metallic atomic-size contacts was crucial to elucidate the
relation between the chemical valence of the atoms and the charge transport
characteristics
\cite{Brandbyge1997,Cuevas1998,Scheer1998,Agrait2003}. Furthermore, it has
been shown that the electronic transmission coefficients of atomic contacts
and molecular junctions can be determined experimentally with the help of
superconductivity \cite{Scheer1998,Scheer1997,Scheer2001,Schirm2013} or by
measuring shot noise
\cite{Brom1999,Cron2001,Djukic2006,Kiguchi2008,Vardimon2013,Karimi2016}.

All this suggests that it would be very interesting to carry out similar
investigations in the case of coherent phonon transport. A related analysis to
those of transmission eigenchannels is that of the mode-dependent
transmission, which can be naturally performed with mode-matching-based
approaches \cite{Tanaka2005,Antonyuk2005,Murphy2007} or with the help of NEGF
techniques \cite{Ong2015,Sadasivam2017}. However, mode-dependent transmission
studies do not actually provide information on the eigenchannels in the
central device part, and they are restricted to bulk systems with
translational symmetry. For this reason such a kind of analysis is not
suitable for atomic and molecular junctions that lack spatial symmetry. Those
atomic-scale systems are better described by means of a combination of ab
initio methods and NEGF techniques \cite{Segal2016}. The problem with
NEGF-based approaches is that they do not provide immediate access to the
scattering states of the system, which makes the determination of meaningful
eigenchannels a challenging task. For this reason, the calculations performed
with NEGF techniques are often interpreted with the help of the local density
of states (LDOS) \cite{Xie2011,Wang2011,Ouyang2012,Peng2014} rather than in
terms of eigenchannels. In the case of electronic transport, Paulsson and
Brandbyge \cite{Paulsson2007} were able to solve this problem and showed how
the eigenchannels can be obtained from information about the subspace of the
central part of the device only, i.e., from data that is readily available in
NEGF-based approaches.  The goal of this paper is to extend those ideas to
obtain the eigenchannels for coherent phonon transport. In particular, we
present here a general procedure to extract these eigenchannels in NEGF-based
calculations. Moreover, we show that this formulation can be combined with
state-of-the-art ab initio methods, and we illustrate this fact with the
analysis of the phonon eigenchannels in a variety of single-atom and
single-molecule junctions of special interest.

The rest of the paper is organized as follows. In Sec.~\ref{sec:Procedure} we
present our procedure to determine transmission eigenchannels for coherent
phonon transport in nanoscale systems. For this purpose, we introduce in
subsection~\ref{sec-SS} the main equations to describe phonon transport, and
we explain how they can be solved formally in terms of scattering states using
NEGFs. In subsection~\ref{sec-SF} we discuss the spectral function, which
plays a key role in the determination of the eigenchannels, and we show how it
is connected to the scattering states of the system. Finally, in
subsection~\ref{sec-TE} we present the procedure to determine the
eigenchannels from a suitably chosen transmission probability matrix using
only information about the subspace of the central part of the device. We
illustrate this method in Sec.~\ref{sec-EX} through a detailed discussion of
examples ranging from a simple toy model, consisting of a one-dimensional (1D)
chain, to various realistic systems such as a gold atomic contact and
single-molecule junctions based on an alkane chain, a C$_{60}$ molecule, and a
benzene derivative. We close the paper in Sec.~\ref{sec-Concl} with a brief
summary of our main conclusions.

\section{Theoretical procedure}\label{sec:Procedure}
In this section, we present the theoretical formalism to determine
transmission eigenchannels for phonons. In analogy to electronic transport
\cite{Paulsson2007,Buerkle2012} we define the eigenchannel with number $\mu$
as particular scattering state that can be computed as the eigenfunction of a
suitably chosen transmission probability matrix, while $\tau_\mu$ is the
corresponding transmission eigenvalue.

\subsection{Scattering states}\label{sec-SS}

We start our analysis of the coherent phonon transport in a given nanoscale
junction with the description of the phononic system in the harmonic
approximation. Within this approximation, the phonons in an infinite spatial 
domain $\Omega$ are described by the following Hamiltonian
\begin{equation}
  \hat{H} = \sum_{i \in \Omega,\alpha} \frac{\hat{p}_{i\alpha}^2}{2} +
  \frac{1}{2\hbar^2}\sum_{i,j\in \Omega,\alpha,\beta} \hat{q}_{i\alpha}
  K_{i\alpha,j\beta} \hat{q}_{j\beta}\label{eq:Hfull}
\end{equation}
Here, $\hat{q}_{i\alpha}=\hat{Q}_{i\alpha}\sqrt{m_i}$ is the mass-weighted
displacement operator of atom $i$ with mass $m_i$,
$\hat{p}_{i\alpha}=\hat{P}_{i\alpha}/\sqrt{m_i}$ is the corresponding
mass-scaled canonical momentum operator, and $K_{i\alpha,j\beta} =
\hbar^2\partial_{i\alpha}\partial_{j\beta}E_{\mathrm{BO}}/\sqrt{m_im_j}$ is
the dynamical matrix, which is the mass-weighted second derivative of the
Born-Oppenheimer energy. Displacements of atoms $i,j$ are assumed to be along
the Cartesian axes $\alpha,\beta=\mathrm{x,y,z}$. The operators in
Eq.~(\ref{eq:Hfull}) fulfill the standard commutation relations
$[\hat{q}_{i\alpha},\hat{q}_{j\beta}] =
[\hat{p}_{i\alpha},\hat{p}_{j\beta}]=0$ and
$[\hat{q}_{i\alpha},\hat{p}_{j\beta}]=\mathrm{i}\hbar
\delta_{i,j}\delta_{\alpha,\beta}$.

In a typical transport setup the domain $\Omega$ is divided into three parts:
a semi-infinite left (L) lead, a finite central (C) part and a semi-infinite
right (R) lead. The Hamilton operator can then be written as
\begin{equation}
  \hat{H} = \hat{H}_\mathrm{L} + \hat{H}_\mathrm{C} + \hat{H}_\mathrm{R},
\end{equation}
with
\begin{eqnarray}
  \hat{H}_X=\sum_{i\in X,\alpha} \frac{\hat{p}_{i\alpha}^2}{2} + 
  \frac{1}{2\hbar^2}\sum_{i\in X,j \in \Omega,\alpha,\beta}\hat{q}_{i\alpha}K_{i\alpha,j\beta}\hat{q}_{j\beta}, 
\end{eqnarray}
where $X$=L,C,R. Since it is customary for phonon transport to work in the
Heisenberg picture, we shall consider the Heisenberg operator
\begin{equation}
  \hat{q}_{i\alpha}(t)=e^{\mathrm{i}\hat{H}t/\hbar}\hat{q}_{i\alpha}e^{-\mathrm{i}\hat{H}t/\hbar}.
\end{equation} 
It is straightforward to show that this operator fulfills the following
equation of motion
\begin{equation} \label{eq:EOM}
  \hbar^2\frac{d^2\hat{q}_{i\alpha}(t)}{dt^2} = -\sum_{j\in \Omega,\beta}
  K_{i\alpha,j\beta}\hat{q}_{j\beta}(t).
\end{equation}
The full solution to this equation of motion is formed from two sets of states
\cite{Farina1973}. One set includes propagating states with a continuous
energy spectrum. It is generated from the electrodes, which we assume to be
perfect semi-infinite crystals without defects. The upper cutoff energy
$E_{\mathrm{c}}$ of the spectrum is determined by the Debye energy of the left
or right electrode material and is set to the maximum of the two values. The
other set is formed by bound states with a discrete energy spectrum,
originating from the finite central region. The bound states are not important
for coherent transport, because they do not contribute to the
transmission. Nevertheless, we take them into account in our considerations,
since they are crucial for the normalization of the states, as we will discuss
below.

The solution of Eq.~(\ref{eq:EOM}) can then be expressed in terms of the
normal modes of the propagating and bound sets as
\begin{eqnarray} \label{eq:solution}
  \hat{q}_{i\alpha}(t) & = & \int_0^{E_{\mathrm{c}}} \mathrm{d}E \sum_m
  \frac{\hbar}{\sqrt{2 E}} \Big( b_m^\dagger(E) \Phi_{m,i\alpha}^*(E)
  e^{\mathrm{i} E t/\hbar} +\mathrm{h.c.} \Big) \nonumber \\ & & + \sum_m
  \frac{\hbar}{\sqrt{2 E_m}} \left( \bar{b}_m^\dagger \bar{\Phi}_{m,i\alpha}^*
  e^{\mathrm{i} E_m t/\hbar} +\mathrm{h.c.}  \right),
\end{eqnarray}
where h.c.\ denotes Hermitian conjugation. The normal mode operators fulfill
standard commutation relations with the only nonvanishing commutators being
$[b_m(E),b_n^\dagger(E')] =\delta_{mn}\delta(E-E')$ and
$[\bar{b}_m,\bar{b}_n^\dagger]=\delta_{mn}$. In these expressions,
$\Phi_{m,i\alpha}(E)$ is the component of the normal mode vector
$\boldsymbol{\Phi}_m(E)$ on atom $i$ for the displacement along $\alpha$,
which solves the following eigenvalue problem
\begin{equation}\label{eq:KPhi-continuous}
  K\boldsymbol{\Phi}_m(E)=E^2 \boldsymbol{\Phi}_m(E)
\end{equation}
for a given energy $E$. Here, $m$ runs over all degenerate states with energy
$E$. Similar relations hold for the bound states, where
$\bar{\boldsymbol{\Phi}}_m$ is the normal mode vector $m$, which solves
\begin{equation}\label{eq:KPhi-discrete}
  K\bar{\boldsymbol{\Phi}}_m=E_m^2 \bar{\boldsymbol{\Phi}}_m.
\end{equation}
In this case, the index $m$ enumerates all bound states. Overall, the normal
mode vectors are normalized such that
\begin{eqnarray}\label{eq:Enorm}
  \int_0^{E_{\mathrm{c}}} \mathrm{d}E \sum_m \boldsymbol{\Phi}_{m,i\alpha}^*(E)
  \boldsymbol{\Phi}_{m,j\beta}(E) & & \nonumber
  \\ +\sum_m\bar{\boldsymbol{\Phi}}_{m,i\alpha}^*
  \bar{\boldsymbol{\Phi}}_{m,j\beta} & = & \delta_{ij}\delta_{\alpha\beta}.
\end{eqnarray} 

Since we are interested in the formulation of transport as a scattering
problem, we solve Eq.~(\ref{eq:KPhi-continuous}) for the propagating set of
states by starting from the solutions of the uncoupled subsystems $X$ and
treat the coupling between the different parts, $K_{XY}$ with $Y\neq X$, as a
perturbation $K_1$. For this reason we write $K= K_0 + K_1$ with
\begin{equation}\label{eq:K0}
  K_0=
  \begin{pmatrix}
    K_{\mathrm{LL}} &0& 0 \\
    0& K_{\mathrm{CC}} & 0 \\
    0& 0 & K_{\mathrm{RR}} 
  \end{pmatrix}
\end{equation}
and
\begin{equation}\label{eq:K1}
  K_1=
  \begin{pmatrix}
    0 & K_{\mathrm{LC}} & 0 \\
    K_{\mathrm{CL}}& 0 & K_{\mathrm{CR}} \\
    0 & K_{\mathrm{RC}} & 0 
  \end{pmatrix}.
\end{equation}
Note that we assume here and henceforth that left and right parts are
decoupled, meaning that $K_{\mathrm{LR}}=K_{\mathrm{RL}}^\dagger=0$. For the
eigenvalue $E^2$, we arrive in this way at a general solution
$\boldsymbol{\Phi}_m(E) =
\left(\boldsymbol{\Phi}_{m,\mathrm{L}}(E),\boldsymbol{\Phi}_{m,\mathrm{C}}(E),
\boldsymbol{\Phi}_{m,\mathrm{R}}(E)\right)^T$, which can be expressed by using
the Green's function formalism as follows
\begin{equation}
  \boldsymbol{\Phi}_{m,X}(E) = \boldsymbol{\varphi}_{m,X}(E) + \sum_{Y\neq
      X} d_{XX}^{\mathrm{r}}(E) K_{XY} \boldsymbol{\Phi}_{m,Y}(E).
  \label{eq:Scat-states-1}
\end{equation}
Here $\boldsymbol{\varphi}_{m,X}(E)$ is the solution of the unperturbed
system, i.e., $\left(E^2 - K_{XX}\right) \boldsymbol{\varphi}_{m,X}(E) = 0$,
and
\begin{equation}
  d_{XX}^{\mathrm{r}}(E)=\left[(E+\mathrm{i}\eta)^2-K_{XX}\right]^{-1}
\end{equation}
is the retarded Green's function of the unperturbed solution with an
infinitesimal parameter $\eta>0$. The states in Eq.~(\ref{eq:Scat-states-1})
can also be written in terms of the retarded Green's function of the full
system
\begin{equation} \label{eq:D}
  D^{\mathrm{r}}(E)=\left[ (E+\mathrm{i}\eta )^2 - K \right]^{-1}
\end{equation}
as
\begin{equation} \label{eq:LS}
\boldsymbol{\Phi}_{m,X}(E) = \boldsymbol{\varphi}_{m,X}(E)
+\sum_Z\sum_{Y\neq Z} D_{XZ}^{\mathrm{r}}(E) K_{ZY}
\boldsymbol{\varphi}_{m,Y}(E).
\end{equation}
From this equation, we define the scattering states
$\boldsymbol{\Phi}_m^{\mathrm{L}}(E)$ [$\boldsymbol{\Phi}_m^{\mathrm{R}}(E)$]
generated from unperturbed states that enter the junction region from the left
[right] lead, which are special solutions with the boundary conditions
$\boldsymbol{\varphi}_{m,\mathrm{C}}(E)=0$ and simultaneously
$\boldsymbol{\varphi}_{m,\mathrm{R}}(E)=0$
[$\boldsymbol{\varphi}_{m,\mathrm{L}}(E)=0$]. We will show in the next section
that, apart from contributions due to bound states, these left- and
right-incoming states give rise to the spectral function of the central part.

\subsection{Spectral function}\label{sec-SF}

The phonon spectral function plays a central role in the determination of the
transmission eigenchannels. This function is given in terms of the phonon
Green's functions as follows
\begin{equation}
  A(E) = \mathrm{i}\{ D^{\mathrm{r}}(E) - \left[ D^{\mathrm{r}}(E)
    \right]^\dagger \} .\label{eq:Afull}
\end{equation}
Making use of the propagating and bound sets of solutions to
Eqs.~(\ref{eq:KPhi-continuous}) and (\ref{eq:KPhi-discrete}), the spectral
function can be rewritten as
\begin{eqnarray} \label{eq:Aw}
  A(E) & = & -2\int_0^{E_{\mathrm{c}}} \mathrm{d}E' \sum_m \mathrm{Im} \left[
    \frac{\boldsymbol{\Phi}_m(E') \boldsymbol{\Phi}_m^\dagger(E')}
         {(E+\mathrm{i}\eta)^2-E'^2} \right] \nonumber\\ & & -2 \sum_m
  \mathrm{Im} \left[ \frac{\bar{\boldsymbol{\Phi}}_m
      \bar{\boldsymbol{\Phi}}_m^\dagger} {(E+\mathrm{i}\eta)^2-E_m^2} \right].
\end{eqnarray}
We note that this form of the spectral function is consistent with the
standard definition of the Green's functions used for the derivation of the
Landauer formula \cite{Mingo2006}. Those Green's functions are defined in
terms of the operators $\hat q_{i\alpha}(t)$ of Eq.~(\ref{eq:solution}), and
such a starting point also leads to Eq.~(\ref{eq:Aw}). Now, using that
$\lim\limits_{\eta\rightarrow 0} \mathrm{Im}
\left[1/(E+\mathrm{i}\eta)\right]=-\pi\delta(E)$, we can express the spectral
function as follows
\begin{eqnarray}\label{eq:Acc}
  A(E) & = & \frac{\pi}{E} \sum_m
  \boldsymbol{\Phi}_m(E)\boldsymbol{\Phi}_m^\dagger(E)\nonumber \\ & &+\sum_m
  \frac{\pi}{E_m}\delta(E-E_m)
  \bar{\boldsymbol{\Phi}}_m\bar{\boldsymbol{\Phi}}_m^\dagger \nonumber\\ &=&
  \frac{\pi}{E}\rho(E),
\end{eqnarray}
where 
\begin{equation}
  \rho(E)=\sum_m
  \boldsymbol{\Phi}_m(E)\boldsymbol{\Phi}_m^\dagger(E)+\sum_m\delta(E-E_m)
  \bar{\boldsymbol{\Phi}}_m\bar{\boldsymbol{\Phi}}_m^\dagger
\end{equation}
is the phonon density of states. This shows that the sets of states
$\boldsymbol{\Phi}_m(E)$ and $\bar{\boldsymbol{\Phi}}_m$, respectively,
contain the information about the density of states at a given energy.

After these general considerations, we now address the spectral function of
our scattering problem. We obtain the retarded Green's function in the central
region from Eq.~(\ref{eq:D}), and it is given by the Dyson equation
\begin{equation}\label{eq:DCC}
  D_{\mathrm{CC}}^{\mathrm{r}}(E)= \left[ (E+\mathrm{i}\eta)^2 - K_{\mathrm{CC}} -
    \Pi_{\mathrm{L}}^{\mathrm{r}}(E)- \Pi_{\mathrm{R}}^{\mathrm{r}}(E) \right]^{-1},
\end{equation}
where $\Pi_Z^{\mathrm{r}}(E) =
K_{\mathrm{C}Z}d_{ZZ}^{\mathrm{r}}(E)K_{Z\mathrm{C}}$ with $Z$=L,R is the
embedding self-energy due to the coupling to the leads. The spectral function
of the central part can then be expressed using Eq.~(\ref{eq:Afull}) as
\begin{widetext}
  \begin{eqnarray}\label{eq:spectral-func}
    A_{\mathrm{C}}(E) & = & \mathrm{i} D_{\mathrm{CC}}^{\mathrm{r}}(E) \left
    \{ \left[ D_{\mathrm{CC}}^{\mathrm{r}}(E)^\dagger \right]^{-1} -
    D_{\mathrm{CC}}^{\mathrm{r}}(E)^{-1} \right \}
    D_{\mathrm{CC}}^{\mathrm{r}}(E)^\dagger \nonumber \\ &=&
    \sum_{Z=\mathrm{L,R}} D_{\mathrm{CC}}^{\mathrm{r}}(E)\Lambda_{Z}(E)
    D_{\mathrm{CC}}^{\mathrm{r}}(E)^\dagger-4\mathrm{i}\eta E
    D_{\mathrm{CC}}^{\mathrm{r}}(E) D_{\mathrm{CC}}^{\mathrm{r}}(E)^\dagger
  \end{eqnarray}
\end{widetext}
with $\Lambda_Z(E) = \mathrm{i} \left[ \Pi_Z^{\mathrm{r}}(E) -
  \Pi_Z^{\mathrm{r}}(E)^\dagger \right] =
K_{\mathrm{C}Z}a_{Z}(E)K_{Z\mathrm{C}}$ and $a_{Z}(E) = (\pi/E) \sum_m
\boldsymbol{\varphi}_{m,Z}(E)\boldsymbol{\varphi}_{m,Z}^\dagger(E)$. While the
last term $-4\mathrm{i}\eta E D_{\mathrm{CC}}^{\mathrm{r}}(E)
D_{\mathrm{CC}}^{\mathrm{r}}(E)^\dagger$ in Eq.~(\ref{eq:spectral-func})
corresponds to the bound-state contributions, one can show that the two terms
$D_{\mathrm{CC}}^{\mathrm{r}}(E)\Lambda_{Z}(E)
D_{\mathrm{CC}}^{\mathrm{r}}(E)^\dagger$ for $Z=\mathrm{L,R}$ are related to
the scattering states $\boldsymbol{\Phi}_m^{\mathrm{L}}(E)$ and
$\boldsymbol{\Phi}_m^{\mathrm{R}}(E)$. This can be demonstrated as follows
\begin{widetext}
  \begin{eqnarray}\label{eq:ACZ}
    A_{\mathrm{C}}^{Z}(E) & = & \frac{\pi}{E}\sum_m P_{\mathrm{C}}
    \boldsymbol{\Phi}_m^{Z}(E) \boldsymbol{\Phi}_m^Z(E)^\dagger P_{\mathrm{C}}
    \nonumber \\ & = & \frac{\pi}{E}\sum_m P_{\mathrm{C}} \left[
      \boldsymbol{\varphi}_{m}^Z(E) + D^{\mathrm{r}}(E) K_1
      \boldsymbol{\varphi}_{m}^Z(E)\right]
    \left[\boldsymbol{\varphi}_{m}^{Z}(E)^\dagger +
      \boldsymbol{\varphi}_{m}^{Z}(E)^\dagger K_1^\dagger
      D^{\mathrm{r}}(E)^\dagger\right] P_{\mathrm{C}} \nonumber\\ & = &
    D_{\mathrm{CC}}^{\mathrm{r}}(E) \Lambda_{Z}(E)
    D_{\mathrm{CC}}^{\mathrm{r}}(E)^\dagger,
  \end{eqnarray}
\end{widetext}
where we have used Eq.~(\ref{eq:LS}) for the scattering states,
$\boldsymbol{\varphi}_m^{\mathrm{L}}(E) =
\left(\boldsymbol{\varphi}_{m,\mathrm{L}}(E),0,0\right)^T$,
$\boldsymbol{\varphi}_m^{\mathrm{R}}(E) =
\left(0,0,\boldsymbol{\varphi}_{m,\mathrm{R}}(E)\right)^T$ and the projection
operator
\begin{equation}\label{eq:PC}
  P_{\mathrm{C}}=\sum_{i \in \mathrm{C},\alpha}
  \boldsymbol{e}_{i\alpha}\boldsymbol{e}_{i\alpha}^\dagger.
\end{equation}
In this expression, $\boldsymbol{e}_{i\alpha}$ is a unit vector of the same
dimension as the $\boldsymbol{\Phi}_m(E)$, and its entries are given by
$e_{i\alpha,j\beta}=\delta_{ij}\delta_{\alpha\beta}$.  We have thus shown that
the spectral function of the central part
$A_{\mathrm{C}}(E)=A_{\mathrm{C}}^{\mathrm{L}}(E)+A_{\mathrm{C}}^{\mathrm{R}}(E)+A_{\mathrm{C}}^{\mathrm{B}}(E)$
consists of two spectral functions $A_{\mathrm{C}}^{\mathrm{L}}(E)$ and
$A_{\mathrm{C}}^{\mathrm{R}}(E)$, which can be attributed to scattering states
$\boldsymbol{\Phi}_m^{\mathrm{L}}(E)$ and
$\boldsymbol{\Phi}_m^{\mathrm{R}}(E)$ that enter the central device region
from the left and right leads, respectively, and a part
$A_{\mathrm{C}}^{\mathrm{B}}(E)$ due to bound states.

\subsection{Transmission eigenchannels}\label{sec-TE}

We are now in the position to finally describe the procedure to determine the
transmission eigenchannels. Let us first recall that we assume that the left
and right parts are decoupled [see Eqs.~(\ref{eq:K0}) and
  (\ref{eq:K1})]. Under these conditions and using the NEGF formalism, one can
show that the phononic heat current is given by a Landauer-like formula that
reads
\cite{Zhang2007,Mingo2003,Wang2006a,Wang2007,Yamamoto2006,Wang2014,Das2012}

\begin{equation}
J(T)= \frac{1}{2\pi\hbar}\int_{0}^\infty \mathrm{d}E E \tau(E)
\left[ n_\mathrm{R}(E,T)- n_\mathrm{L}(E,T) \right] ,
\end{equation}
where 
\begin{equation}
  \tau(E) = \mathrm{Tr}\left[D_{\mathrm{CC}}^{\mathrm{r}}(E)
    \Lambda_{\mathrm{L}}(E) D_{\mathrm{CC}}^{\mathrm{r}}(E)^\dagger
    \Lambda_{\mathrm{R}}(E) \right],\label{eq:tauE}
\end{equation}
is the total phonon transmission and $n_Z(E,T)=1/ \{ \exp[ E/(k_{\rm
    B}T_Z)] - 1 \}$ is the Bose-Einstein distribution 
function.

In order to obtain eigenchannels as linear combinations of projections of
scattering states onto the central junction part simultaneously with the
corresponding transmission eigenvalues, we express the transmission using
Eq.~(\ref{eq:ACZ}) with $Z=\mathrm{L}$ as
\begin{eqnarray}
  \tau(E) &=& \mathrm{Tr}\left[
    A_{\mathrm{C}}^{\mathrm{L}}(E)\Lambda_{\mathrm{R}}(E)\right] \nonumber
  \\ &=& \frac{\pi}{E} \sum_m \boldsymbol{\Phi}_m^{\mathrm{L}}(E)^\dagger
  P_{\mathrm{C}} \Lambda_{\mathrm{R}}(E) P_{\mathrm{C}}
  \boldsymbol{\Phi}_m^{\mathrm{L}}(E).\label{eq:tauE-ACL-LambdaR}
\end{eqnarray}
Inspired by this expression, we define the transmission probability matrix
\begin{equation}
  \tau_{mn}^{(1)}(E) = \frac{\pi}{E} \boldsymbol{\Phi}_m^{\mathrm{L}}(E)^\dagger P_{\mathrm{C}}\Lambda_{\mathrm{R}}(E)
  P_{\mathrm{C}}\boldsymbol{\Phi}_{n}^{\mathrm{L}}(E),\label{eq:tau1mn-initial}
\end{equation}
which is actually the matrix that we shall diagonalize to obtain the eigenchannels.

In order to diagonalize this transmission matrix, we follow the procedure for
the electronic problem, as described in Refs.~\cite{Paulsson2007,Buerkle2012},
and perform a spectral decomposition for the central part of the spectral
function
\begin{eqnarray}
  A_{\mathrm{C}}^{\mathrm{L}}(E) &=& \sum_m \tilde{\boldsymbol{\chi}}_m(E) \lambda_m(E)
  \tilde{\boldsymbol{\chi}}_m^\dagger(E) \nonumber \\ &=& \frac{\pi}{E} \sum_m
  \tilde{\boldsymbol{\xi}}_m(E)
  \tilde{\boldsymbol{\xi}}_m^\dagger(E).\label{eq:ACL}
\end{eqnarray}
Here, $\tilde{\boldsymbol{\xi}}_m(E) =
\sqrt{E\lambda_m(E)/\pi}\tilde{\boldsymbol{\chi}}_m(E) $ and
$\tilde{\boldsymbol{\chi}}_m^\dagger(E)\tilde{\boldsymbol{\chi}}_{n}(E) =\delta_{mn}$. As can be
seen from a comparison of Eqs.~(\ref{eq:ACZ}) and (\ref{eq:ACL}), the vectors
$\tilde{\boldsymbol{\xi}}_m(E)= P_{\mathrm{C}}
\boldsymbol{\Phi}_m^{\mathrm{L}}(E)$ originate from the scattering states that
arrive from the left lead via projections onto the central part and are
therefore normalized through the $\boldsymbol{\Phi}_m^{\mathrm{L}}(E)$ [see
  also Eq.~(\ref{eq:Enorm})]. Then, we transform
$\pi\Lambda_{\mathrm{R}}(E)/E$ into the new basis of the
$\tilde{\boldsymbol{\xi}}_m(E)$ through
\begin{eqnarray}
  \tau_{mn}^{(1)}(E)&=& \frac{\pi}{E} \tilde{\boldsymbol{\xi}}_{m}^\dagger(E)
  \Lambda_\mathrm{R}(E) \tilde{\boldsymbol{\xi}}_{n}(E) \nonumber
  \\ &=&\frac{\pi}{E}\left[\tilde{U}^{\dagger}(E) \Lambda_{\mathrm{R}}(E)
    \tilde{U}(E)\right]_{mn},\label{eq:tau1mn}
\end{eqnarray}
where $\tilde{U}(E)=\left(
\tilde{\boldsymbol{\xi}}_1(E),\dots,\tilde{\boldsymbol{\xi}}_{3N_{\mathrm{C}}}(E)\right)$
and $N_{\mathrm{C}}$ is the number of atoms in the central part. The
eigenvectors are solutions of the equation
\begin{equation}\label{eq:diag}
  \sum_n \tau_{mn}^{(1)}(E)c_{n\mu}(E) = \tau_\mu(E) c_{m\mu}(E)
\end{equation}
with $\sum_m c_{m\mu}^*(E)c_{m\nu}(E)=\delta_{\mu\nu}$, and the eigenchannel
$\mu$ in the central region is given by
\begin{eqnarray}\label{eq:back-transform}
  \tilde{\boldsymbol{\Psi}}_\mu(E)&=&\sum_{m}c_{m\mu}(E)\tilde{\boldsymbol{\xi}}_{m}(E)\nonumber
  \\
  &=&\sum_{i\in \mathrm{C},\alpha}a_{i\alpha,\mu}(E)\boldsymbol{e}_{i\alpha}.
\end{eqnarray}
with $a_{i\alpha,\mu}(E)=\sum_m \tilde{U}_{i\alpha,m}(E)c_{m\mu}(E)$. The
eigenchannels thus arise from a unitary transformation of the states
$\tilde{\boldsymbol{\xi}}_{m}(E)$.

Let us note that the eigenchannels of Eq.~(\ref{eq:back-transform}) are right
eigenvectors of the transmission probability matrix
$\tau^{(2)}(E)=A_{\mathrm{C}}^{\mathrm{L}}(E)\Lambda_{\mathrm{R}}(E)$ that
appears in the trace of Eq.~(\ref{eq:tauE-ACL-LambdaR}), i.e.,
\begin{equation}
  \tau^{(2)}(E)\tilde{\boldsymbol{\Psi}}_\mu(E)=\tau_\mu(E)\tilde{\boldsymbol{\Psi}}_\mu(E).\label{eq:tau2-eigv}
\end{equation}
This is evident, if the relations in
Eqs.~(\ref{eq:ACL})--(\ref{eq:back-transform}) are used. It is worth pointing
out that apart from $\tau^{(1)}(E)$ or $\tau^{(2)}(E)$, one could eventually
consider other forms for the transmission probability matrix. For instance, we
might want to use $\tau^{(3)}(E)=t(E) t^\dagger(E)$ with
$t(E)=\Lambda_{\mathrm{R}}^{1/2}(E)D_{\mathrm{CC}}^{\mathrm{r}}(E)\Lambda_{\mathrm{L}}^{1/2}(E)$. Given
an eigenchannel $\tilde{\boldsymbol{\Psi}}_\mu(E)$ with eigenvalue
$\tau_\mu(E)$ of $\tau^{(2)}(E)$ [see Eq.~(\ref{eq:tau2-eigv})], we find that
$\Lambda_{\mathrm{R}}^{1/2}(E)\tilde{\boldsymbol{\Psi}}_\mu(E)$ is an
eigenvector of $\tau^{(3)}(E)$ with the same eigenvalue $\tau_\mu(E)$. Similar
to the electronic case \cite{Buerkle2012}, we thus observe that the
eigenvectors of $\tau^{(3)}(E)$ do no longer result from a unitary
transformation of scattering states that are projected onto the center via
$P_{\mathrm{C}}$ [see Eq.~(\ref{eq:PC})], as it was the case when using
$\tau^{(2)}(E)$ [see Eq.~(\ref{eq:back-transform})]. Instead, the matrix
$\Lambda_{\mathrm{R}}^{1/2}(E)$ destroys simultaneously the $P_{\mathrm{C}}$
projection property as well as the normalization [see Eq.~(\ref{eq:Enorm})],
and a comparison of the amplitudes of eigenchannels of $\tau^{(3)}(E)$ would
thus not be meaningful.

As it is obvious from the relation $\tilde{\boldsymbol{\xi}}_m(E)=
P_{\mathrm{C}} \boldsymbol{\Phi}_m^{\mathrm{L}}(E)$,
Eqs.~(\ref{eq:tau1mn-initial})--(\ref{eq:back-transform}) yield left-incoming
eigenchannels originating from the scattering states
$\boldsymbol{\Phi}_m^{\mathrm{L}}(E)$. This means that the lattice vibrations
arrive at the scattering region from the left lead and are subsequently
transmitted to the right lead or scattered back to the left one. In order to
obtain right-incoming eigenchannels, it is sufficient to start from
$\tau_{mn}^{(4)}(E) = \pi \boldsymbol{\Phi}_m^{\mathrm{R}}(E)^\dagger
P_{\mathrm{C}}\Lambda_{\mathrm{L}}(E)
P_{\mathrm{C}}\boldsymbol{\Phi}_{n}^{\mathrm{R}}(E)/E$ in
Eq.~(\ref{eq:tau1mn-initial}) or
$\tau^{(5)}(E)=A_{\mathrm{C}}^{\mathrm{R}}(E)\Lambda_{\mathrm{L}}(E)$ in
Eq.~(\ref{eq:tau2-eigv}). The corresponding transmission probability matrices
are obtained by rearranging the expression in the trace of Eq.~(\ref{eq:tauE})
through cyclic permutation, by exploiting the definition of
$A_{\mathrm{C}}^{\mathrm{R}}(E)$ in Eq.~(\ref{eq:ACZ}), and by noting that it
can also be written in the form
$A_{\mathrm{C}}^{\mathrm{R}}(E)=D_{\mathrm{CC}}^{\mathrm{r}}(E)^\dagger
\Lambda_{Z}(E) D_{\mathrm{CC}}^{\mathrm{r}}(E)$ through the relations given in
Eqs.~(\ref{eq:Afull}) and (\ref{eq:spectral-func}).

The eigenchannels in the complete system space
$\boldsymbol{\Psi}_\mu(E)=\sum_{m}c_{m\mu}(E)\boldsymbol{\Phi}_{m}^{\mathrm{L}}(E)$
can be obtained from the $\tilde{\boldsymbol{\Psi}}_\mu(E)$ in
Eq.~(\ref{eq:back-transform}) by omitting the projection $P_{\mathrm{C}}$ on
the central device part. We will however focus in the following on
device-projected eigenchannels. The $\tilde{\boldsymbol{\Psi}}_\mu(E)$ are
normalized according to Eq.~(\ref{eq:Enorm}), because they are constructed
through a unitary transformation with the $c_{m\mu}$ from the
$\tilde{\boldsymbol{\xi}}_{m}(E)$.  Consequently, they are measured in units
of J$^{-1/2}$.  There is also a global phase factor that needs to be fixed for
every eigenchannel $\tilde{\boldsymbol{\Psi}}_\mu(E)$. In the examples shown
below, we will simply set the component of a certain atom to a real value for
the one-dimensional chain. In the ab initio calculations the numerical
routines used for computing the eigenvectors determine the phase factor, which
may thus vary both with $E$ and $\mu$.

We want to transform now the $\tilde{\boldsymbol{\Psi}}_{m}(E)$ to
displacement vectors measured in units of m, in analogy to what is done when
normal modes of finite systems are calculated classically from the eigenvalue
equation~(\ref{eq:KPhi-discrete}). For this reason, we divide
$\tilde{\boldsymbol{\Psi}}_\mu(E)$ by $\sqrt{m_i}$ [see also the mass factor
  in Eq.~(\ref{eq:solution})] and multiply in addition with an
energy-dependent scaling factor $s(E)$ of unit J$^{1/2}$m. In this way the
complex displacements of the central part of the eigenchannels are obtained as
\begin{eqnarray}
  \tilde{\boldsymbol{Q}}_\mu(E) &=& \sum_{i \in \mathrm{C},\alpha}
  \frac{s(E)}{\sqrt{m_i}}a_{i\alpha,\mu}(E) \boldsymbol{e}_{i\alpha}
  \nonumber\\ &=& \sum_{i \in \mathrm{C},\alpha} \frac{s(E)}{\sqrt{m_i}}
  |a_{i\alpha,\mu}(E)| e^{\mathrm{i}\theta_{i\alpha,\mu}(E)}
  \boldsymbol{e}_{i\alpha}.
  \label{eq:QE}
\end{eqnarray}

Equation~(\ref{eq:QE}) shows that each atomic displacement acquires a phase
factor due to the incident wave from the left lead.  Note that the
displacements of the eigenchannels $\tilde{\boldsymbol{Q}}_\mu(E)$ in
Eq.~(\ref{eq:QE}) are proportional to the eigenchannels
$\tilde{\boldsymbol{\Psi}}_\mu(E)$ in Eq.~(\ref{eq:back-transform}), if all of
the $m_i$ are the same, as it is the case in monoatomic junctions. In
contrast, the proportionality is broken for hetero-atomic junctions. We have
furthermore introduced a real-valued scaling factor $s(E)$ in
Eq.~(\ref{eq:QE}), which we may adjust for an optimized visualization of
displacements at each energy $E$. In this way, eigenchannel displacements
$\tilde{\boldsymbol{Q}}_\mu(E)$ at different energies should only be compared
on qualitative grounds, while they are fully comparable at a certain fixed
energy.

The full solution for a wave moving from left to right at an energy $E$ is
\begin{equation}\label{eq:QtE}
  \tilde{\boldsymbol{Q}}_{\mu}(t,E) = \tilde{\boldsymbol{Q}}_{\mu}(E)e^{-\mathrm{i} E t/\hbar}.
\end{equation}
Obviously,
$\tilde{\boldsymbol{Q}}_{\mu}(t=0,E)=\tilde{\boldsymbol{Q}}_{\mu}(E)$. The
time dependence of the real part of the eigenchannel displacement vector
$\mathrm{Re}\tilde{\boldsymbol{Q}}_{\mu}(t,E)$ can be shown in a movie, and we
refer the reader to the supplemental material for examples \cite{SM}, which
will be discussed in the next section. However, for illustrative purposes we
shall often restrict ourselves in the following to the representation of the
real part of the eigenchannel displacements at time $t=0$, i.e.,
$\mathrm{Re}\tilde{\boldsymbol{Q}}_{\mu}(t=0,E)=\mathrm{Re}\tilde{\boldsymbol{Q}}_{\mu}(E)$.

\section{Examples}\label{sec-EX}

We apply now the procedure described in the previous section to determine the
phonon eigenchannels in different situations. The examples range from a
one-dimensional chain, which can be solved analytically, to fully numerical
cases of atomic and molecular junctions in three dimensions. The systems have
been selected to show the versatility of the method, which is applicable to
any system exhibiting phase-coherent phonon transport.

Let us also point out that in all cases studied below, we only present
the results for the left-incoming eigenchannels, since the junctions studied
are rather symmetric. The right-incoming eigenchannels show a similar
behavior and can be obtained at the same computational cost in an analogous
procedure, as explained above.

\subsection{1D chain}\label{sec:1D-chain}

\begin{figure}
\includegraphics[width=1.0\columnwidth]{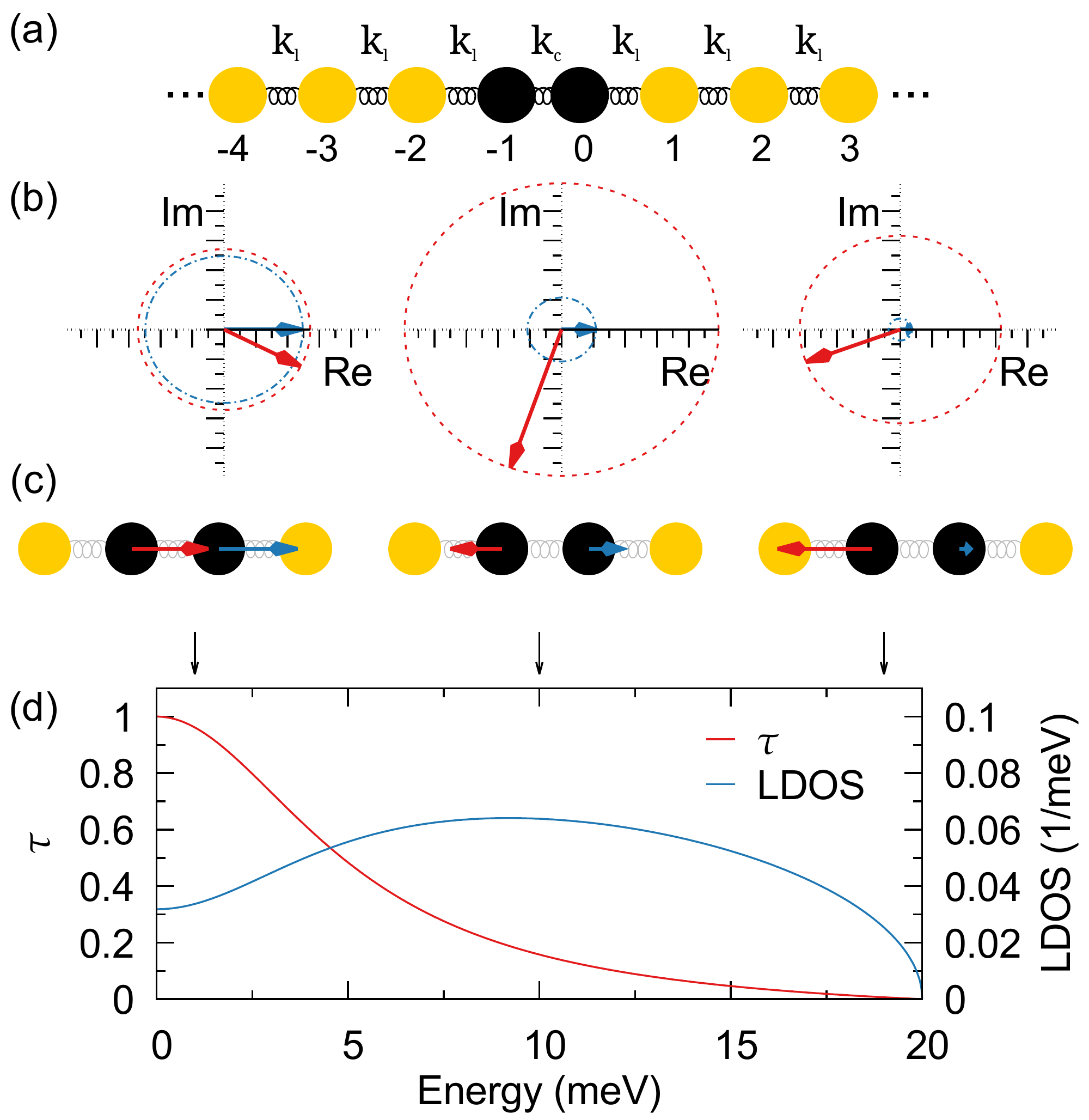}
\caption{(a) Sketch of the 1D junction. Two semi-infinite leads with
  nearest-neighbor coupling constants $k_{\mathrm{l}}$ are connected at sites
  $-1$ and $0$ with the coupling constant $k_{\mathrm{c}}$. The region
  $(-\infty,-2]$ is considered as the L part, the region $[-1,0]$ as the C
part and $[1,\infty)$ as the R part. (b) Transmission eigenchannel
  $\tilde{\boldsymbol{Q}}_1(E,t=0)$ for the 1D chain represented in the
  complex plane for energies of $E=1$~meV, 10~meV, and 19~meV. The red arrow
  shows the complex number $\tilde{Q}_{1,-1}(E,t=0)$ for the atom
  $-1$, the blue one $\tilde{Q}_{1,0}(E,t=0)$ for atom $0$. (c)
  The same as in panel (b), but now we display only the real part of the
  solution $\mathrm{Re}\tilde{\boldsymbol{Q}}_1(E,t=0)$. (d) The corresponding
  transmission eigenvalue $\tau_1(E)$ as a function of energy together with
  the LDOS of one of the atoms ($-1$ or $0$) in the central part. The energies
  of those eigenchannels, which are studied in panels (b) and (c), are
  indicated with arrows. For panels (b)--(d), we assumed spring constants of
  $k_{\mathrm{l}}=100$~meV$^2$ and $k_{\mathrm{c}}=20$~meV$^2$. }
\label{fig:1d}
\end{figure}

We now consider the case of a 1D atomic chain, where the whole procedure for
the determination of phonon transmission eigenchannels can be carried out
analytically. The system that we are interested in is depicted in
Fig.~\ref{fig:1d}(a). In this model junction the C part consists of two atoms,
labeled $-1$ and $0$ and colored in black. These two atoms are coupled through
a spring with force constant $k_{\mathrm{c}}$. The leads are described by two
semi-infinite chains of coupled harmonic oscillators with nearest-neighbor
coupling constant $k_{\mathrm{l}}$. The left (right) lead is connected to atom
$-1$ ($0$) in the central region with a coupling constant
$k_{\mathrm{l}}$. Since the atomic movements are assumed to happen along the
direction of the chain, $\alpha$ reduces to a single component, and the
compound index $i\alpha$ simplifies to just the atom index $i$ in the
following. Furthermore we assume that all atoms in the L, C and R parts have
the same mass $m_i=m$.

The Green's function of the central part $D_{\mathrm{CC}}^{\mathrm{r}}(E)$ can
be obtained from Eq.~(\ref{eq:DCC}) using
\begin{equation}\label{eq:KCC-1D}
  K_{\mathrm{CC}} = \begin{pmatrix}
    k_{\mathrm{c}}+k_{\mathrm{l}} &-k_{\mathrm{c}}\\
    -k_{\mathrm{c}} &k_{\mathrm{c}}+k_{\mathrm{l}}
  \end{pmatrix}
\end{equation}
together with the self-energies
\begin{equation}
  \Pi_{\mathrm{L}}^{\mathrm{r}}(E) = f(E) 
  \begin{pmatrix}
    1 & 0 \\
    0 & 0
  \end{pmatrix}, ~~
  \Pi_{\mathrm{R}}^{\mathrm{r}}(E) = f(E) 
  \begin{pmatrix}
    0 & 0 \\
    0 & 1
  \end{pmatrix} ,
\end{equation}
where $f(E) = (E^2 - 2k_{\mathrm{l}} - E \sqrt{E^2 - 4k_{\mathrm{l}}})/2$. Thus, the
corresponding linewidth-broadening matrices can be written as
\begin{equation}
  \Lambda_{\mathrm{L}}(E) = g(E) 
  \begin{pmatrix}
    1 & 0 \\
    0 & 0
  \end{pmatrix}, ~~ 
  \Lambda_{\mathrm{R}}(E) = g(E) 
  \begin{pmatrix}
    0 & 0 \\
    0 & 1
  \end{pmatrix} ,
\end{equation}
with 
\begin{equation}
  g(E) = \begin{cases}
    E\sqrt{4k_{\mathrm{l}}-E^2} & \mbox{if} ~ E^2 < 4k_{\mathrm{l}}, \\
    0& \mbox{if} ~ E^2 \ge 4k_{\mathrm{l}} .
  \end{cases} 
\end{equation}
From these expressions, the spectral function $A_{\mathrm{C}}^{\mathrm{L}}(E)$
in Eqs.~(\ref{eq:ACZ}) and (\ref{eq:ACL}) is computed. For $E^2 <
4k_{\mathrm{l}}$, the eigenvalues of this matrix are given by
\begin{equation}
  \lambda_1(E)=\frac{\sqrt{ 4 k_{\mathrm{l}}-E^2} [2 k_{\mathrm{c}}^2 + E^2
      (k_{\mathrm{l}}-k_{\mathrm{c}})]} {k_{\mathrm{l}}E [4 k_{\mathrm{c}}^2 +
      E^2 (k_{\mathrm{l}}-2 k_{\mathrm{c}})]}, ~~ \lambda_2(E)=0
\end{equation}
with the corresponding eigenvectors
\begin{eqnarray}
  \tilde{\boldsymbol{\chi}}_1(E) & = & \begin{pmatrix}
    \frac{-E^2 + 2 k_{\mathrm{c}} - \mathrm{i} E \sqrt{4 k_{\mathrm{l}}-E^2
    }}{\sqrt{8k_{\mathrm{c}}^2+4E^2\left(k_{\mathrm{l}}-k_{\mathrm{c}}\right)}} \nonumber \\
    \frac{k_{\mathrm{c}}}{\sqrt{2k_{\mathrm{c}}^2+E^2\left(k_{\mathrm{l}}-k_{\mathrm{c}}\right)}}
  \end{pmatrix},\\
  \tilde{\boldsymbol{\chi}}_2(E) & = & \begin{pmatrix} \frac{k_c (E^2 - 2 k_c + 
  i E \sqrt{4 k_l-E^2 })}{2 \sqrt{(k_c^2 + 
  E^2 (k_l - k_c)) (2 k_c^2 + E^2 (k_l - k_c))}}\\
    \frac{\sqrt{k_c^2 + E^2 (k_l - k_c)}}{\sqrt{2k_c^2 + E^2 (k_l - k_c)}}
  \end{pmatrix},
\end{eqnarray}
which are orthonormal, i.e.,
$\tilde{\boldsymbol{\chi}}_m^\dagger(E)\tilde{\boldsymbol{\chi}}_n(E)=\delta_{mn}$. From
the $\tilde{\boldsymbol{\chi}}_m(E)$ we obtain the C projections of
left-incoming scattering states $\tilde{\boldsymbol{\xi}}_m(E)$ by multiplying
with $\sqrt{E\lambda_m(E)/\pi}$ [see the discussion of
  Eq.~(\ref{eq:ACL})]. Constructing $\tilde{U}(E)=\left(
\tilde{\boldsymbol{\xi}}_1(E),\tilde{\boldsymbol{\xi}}_2(E) \right)$, we
determine $\tau^{(1)}(E)$ via Eq.~(\ref{eq:tau1mn}).  Diagonalizing the
resulting transmission probability matrix [see Eq.~(\ref{eq:diag})], we obtain
the transmission eigenvalues
\begin{equation}
  \tau_1(E)=\frac{k_{\mathrm{c}}^2 (4 k_{\mathrm{l}}-E^2 )}{k_{\mathrm{l}} [4
      k_{\mathrm{c}}^2 + E^2 (k_{\mathrm{l}} -2 k_{\mathrm{c}})]}, ~~
  \tau_2(E)=0
\end{equation}
and eigenvectors
\begin{equation}
  \boldsymbol{c}_1(E)=\begin{pmatrix}
  1\\0
  \end{pmatrix}, ~~
  \boldsymbol{c}_2(E)=\begin{pmatrix}
  0\\1
  \end{pmatrix}.
\end{equation}
These coefficients determine the eigenchannels
$\tilde{\boldsymbol{\Psi}}_\mu(E)$ via Eq.~(\ref{eq:back-transform}). The
time-dependent eigenchannel displacements can now be computed through
Eqs.~(\ref{eq:QE}) and (\ref{eq:QtE}) by transforming the eigenchannels to the
eigenchannel displacement vectors and by multiplying with a time-dependent
phase factor.

Since we assume that the masses of all atoms $m_i=m$ are identical in the 1D
chain, eigenchannels and eigenchannel displacements are proportional
$\tilde{\boldsymbol{Q}}_\mu(E)=s(E)\tilde{\boldsymbol{\Psi}}_\mu(E)/\sqrt{m}$
to each other. We therefore define
$\tilde{\boldsymbol{\Psi}}_\mu(E,t)=\sqrt{m}\tilde{\boldsymbol{Q}}_\mu(t,E)/s(E)$
and use both quantities interchangeably. Choosing the global phase factor of
the eigenchannel such that the component of the atom $0$ is real and positive
at $t=0$, the time-dependent eigenchannels read
\begin{widetext}
  \begin{equation}
    \tilde{\boldsymbol{\Psi}}_1(t,E) = \sqrt{\frac{\sqrt{4 k_{\mathrm{l}}-E^2
        } k_{\mathrm{c}}^2 } {\pi
        k_{\mathrm{l}}\left(E^2(k_{\mathrm{l}}-2k_{\mathrm{c}}) +
        4k_{\mathrm{c}}^2\right)}}\begin{pmatrix} \frac{ -E^2 + 2
        k_{\mathrm{c}} - \mathrm{i}E \sqrt{4 k_{\mathrm{l}}-E^2 }}{2
        k_{\mathrm{c}}} \\ 1
    \end{pmatrix}e^{-\mathrm{i} E t/\hbar}, ~~   \tilde{\boldsymbol{\Psi}}_2(t,E) = 0.\label{eq:Q1tE-1D}
  \end{equation}
\end{widetext}

Let us discuss several points at this stage. We note that there is only a
single eigenchannel with nonvanishing transmission. This is due to the fact
that in our 1D model there is only nearest-neighbor coupling.  Displacements,
which we assume to be along the chain direction, thus need to spread
sequentially from atom to atom. The leads provide a cutoff energy of
$E_{\mathrm{c}}=2\sqrt{k_{\mathrm{l}}}$, above which no propagating states
exist. If $0\leq k_{\mathrm{c}}\leq k_{\mathrm{l}}$ the whole junction shows no
bound states, while they arise if $k_{\mathrm{c}}>k_{\mathrm{l}}\geq0$. Due to the
particular left-right symmetry of our problem, the following relations hold
for $k_{\mathrm{c}}\leq k_{\mathrm{l}}$:
$\rho_{ii}(E)=E[A_{ii}^{\mathrm{L}}(E)+A_{ii}^{\mathrm{R}}(E)]/\pi=
E\mathrm{Tr}[A_{\mathrm{C}}^{\mathrm{L}}(E)]/\pi=E\lambda_1(E)/\pi=|\tilde{\boldsymbol{\Psi}}_1(E)|^2$
with $i=-1,0$. The expressions imply that the square of the norm of the
transmission eigenchannel $1$ follows the LDOS of one of the atoms in the C
part. Integration yields $\int_0^{E_{\mathrm{c}}} \mathrm{d}E
\rho_{ii}(E)=\int_0^{E_{\mathrm{c}}}
\mathrm{d}E|\tilde{\boldsymbol{\Psi}}_1(E)|^2=1$, which is consistent with the
normalization condition in Eq.~(\ref{eq:Enorm}), since there are no bound
states present. For the case $k_{\mathrm{c}}>k_{\mathrm{l}}$, we get
$\int_0^{E_{\mathrm{c}}} \mathrm{d}E|\tilde{\boldsymbol{\Psi}}_1(E)|^2<1$, and
bound-state contributions need to be taken into account in the C part to
fulfill the normalization condition in Eq.~(\ref{eq:Enorm}).

If we now consider the perfect chain with $k_{\mathrm{c}} = k_{\mathrm{l}}$,
the previous results reduce to
\begin{equation}
\tau_1(E)=1, ~~ \tau_2(E)=0
\end{equation}
with
\begin{equation}\label{eq:Q1tE-1Dperfect}
  \tilde{\boldsymbol{\Psi}}_1(t,E) = \frac{1}{\sqrt{\pi \sqrt{4 k_{\mathrm{l}}-E^2 } }}
  \begin{pmatrix}
    u_{-1}(t,E)\\u_0(t,E)
  \end{pmatrix}, ~~ \tilde{\boldsymbol{\Psi}}_2(t,E)  = 0,
\end{equation}
where $u_n(t,E) = \exp[\mathrm{i}k(E)nd-\mathrm{i}E t/\hbar]$. The $u_n(t,E)$
appear as solutions for the equation of motion of atoms arranged in an
infinite chain and coupled by the same nearest-neighbor spring constants
\cite{Ashcroft1976}. Here, we have introduced the wavevector $k(E) = (2/d)
\sin^{-1} (E/E_{\mathrm{c}})$ and neighboring atoms are assumed to be
separated by the distance $d$. Note that the interatomic distance $d$ is not
relevant for our transport problem, which is entirely determined by the force
constant matrix $K$ [see Eqs.~(\ref{eq:K0}) and (\ref{eq:K1})], where force
constants will of course be functions of interatomic distances in realistic
systems. As discussed in the previous paragraph, we find that $\rho_{ii}(E) =
|\tilde{\boldsymbol{\Psi}}_1(E)|^2=2/(\pi\sqrt{4k_{\mathrm{l}}-E^2})$ with
$\int_0^{E_{\mathrm{c}}} \mathrm{d}E \rho_{ii}(E) = 1$.

We want to use now the analytical expressions to examine different
representations of the transmission eigenchannel displacements. For this
purpose we choose the global scaling factor $s/\sqrt{m}$ in Eq.~(\ref{eq:QE})
to be real, energy-independent and of units J$^{1/2}$m/kg$^{1/2}$. In
Fig.~\ref{fig:1d}(b)--(d) we study the transmission, LDOS and eigenchannel
displacements for the 1D chain with $k_{\mathrm{l}}=100$~meV$^2$ and
$k_{\mathrm{c}}=20$~meV$^2$, i.e., in the situation where there are only
propagating states in the junction system. The transmission $\tau(E)$ in
Fig.~\ref{fig:1d}(d) shows a monotonically decreasing behavior with increasing
energy and vanishes above the cutoff energy of $E_{\mathrm{c}}=20$~meV. At the
same time, we plot the LDOS $\rho_{ii}(E)$ of the atom $i=-1,0$ in the central
part, which starts from a finite value at $E=0$, increases to a maximum around
$9$~meV and drops to zero beyond $E_{\mathrm{c}}$. The transmission
eigenchannel displacements $\tilde{\boldsymbol{Q}}_1(t=0,E)$ are
shown in Fig.~\ref{fig:1d}(b) for the energies $E=1$~meV, 10~meV, and 19~meV,
indicated by arrows in Fig.~\ref{fig:1d}(d). The two complex components are
indicated by two arrows in the complex plane. Notice that while the norm of
the eigenvector $\tilde{\boldsymbol{Q}}_1(t,E)$ is proportional to
$\sqrt{\rho_{ii}(E)}$ for the energy-independent $s$ chosen here, the relative
magnitude at the atom $i=0$ as compared to the atom $i=-1$, i.e.,
$|\tilde{Q}_{1,0}(t,E)|/|\tilde{Q}_{1,-1}(t,E)|$, decreases with increasing
energy because a larger portion of the left-incoming wave gets reflected at
the constriction. We also note that the phase difference
$\theta_{0,1}(E)-\theta_{-1,1}(E)$ [see Eq.~(\ref{eq:QE})] between the two
components increases from $0$ at $E=0$ to $\pi$ at $E=20$~meV. With increasing
time the arrows precess around the origin at a constant angular velocity of
$\omega=E/\hbar$, spanning the circle indicated by the dashed lines in the
plot. Since the two atoms typically do not swing in phase, the real parts of
$\tilde{Q}_{1,-1}(t,E)$ and $\tilde{Q}_{1,0}(t,E)$ take maximum amplitudes at
different times.

In Fig.~\ref{fig:1d}(c) we present another way to visualize the eigenchannel
displacements by simply plotting $\mathrm{Re}\tilde{Q}_{1,-1}(t,E)$ and
$\mathrm{Re}\tilde{Q}_{1,0}(t,E)$ at $t=0$ as arrows attached to the
respective atoms. This is actually the representation that we will use in all
the figures shown in the rest of the paper. Notice that due to our choice of
the global phase factor, we get $\theta_{0,1}(E)=0$ and
$\mathrm{Re}\tilde{Q}_{1,0}(t=0,E)$ is hence maximal at $t=0$. In contrast
$\mathrm{Re}\tilde{Q}_{1,-1}(t,E)$ depends both on the absolute value
$|\tilde{Q}_{1,-1}(E)|$ and the phase $\theta_{-1,1}(E)$, as it is visible
from Fig.~\ref{fig:1d}(b). Despite the large $|\tilde{Q}_{1,-1}(E)|$ at
$E=10$~meV, $\mathrm{Re}\tilde{Q}_{1,-1}(t,E)$ is rather small, because
$\theta_{-1,1}(E)\approx-0.6\pi$. In spite of such shortcomings, one gets
an impression of the nature of the atomic motions involved in the
eigenchannel. Indeed, we observe that the eigenchannel displacements at low
energy $E=1$~meV resemble a translational mode of the two atoms, while they
are basically vibrating against each other at $19$~meV, as it is clear from
the evolution of the phase difference $\theta_{0,1}(E)-\theta_{-1,1}(E)$ with
energy, discussed in the previous paragraph. Videos could be used to examine
the full time-dependent dynamics of
$\mathrm{Re}\tilde{\boldsymbol{Q}}_1(t,E)$, but we refrain from this here,
since the simple 1D case is well characterized with the help of
Fig.~\ref{fig:1d}(b).

\subsection{Ab initio results}

After illustrating the method with the simple 1D model, we apply it now to
realistic systems. In these systems, we determine the force constant matrix
for a particular junction geometry with the help of density functional theory
(DFT) and describe the coherent phonon transport within the NEGF formalism
explained in Sec.~\ref{sec:Procedure}. In particular, we will present
different examples of the analysis of the phonon eigenchannels in nanoscale
systems that include a gold single-atom contact \cite{Kloeckner2017a} and
several single-molecule junctions made of gold electrodes that are bridged by
an alkane chain \cite{Kloeckner2016}, a C$_{60}$ molecule
\cite{Kloeckner2017}, and a benzene ring with a bromine substituent
\cite{Kloeckner2017b}, where destructive interference effects show up in the
latter case. Let us stress that we have already studied in detail the phononic
thermal conductance in these systems in the references cited above. Here, we
shall focus on the new insight provided by the analysis of the eigenchannels,
and we refer the reader to those publications for the technical details on the
calculations of the transmission functions.

In our junctions with gold electrodes, the Debye energy of the metal of around
20~meV represents the cutoff energy for the propagating states of the
scattering problem. Because gas-phase molecules typically show vibrations with
energies much above $E_{\mathrm{c}}$, this leads to bound states in the
molecular junctions. They need to be considered for a proper normalization of
the eigenchannels $\tilde{\boldsymbol{\Psi}}_\mu(E)$ in
Eq.~(\ref{eq:back-transform}).

We visualize eigenchannels in all the figures below in terms of the static
picture of the real part of the eigenchannel displacements at $t=0$, i.e.,
$\mathrm{Re}\tilde{\boldsymbol{Q}}_\mu(t=0,E)=\mathrm{Re}\tilde{\boldsymbol{Q}}_\mu(E)$
[see Eqs.~(\ref{eq:QE}) and (\ref{eq:QtE})], but we illustrate the real part
of the full time-dependent solutions
$\mathrm{Re}\tilde{\boldsymbol{Q}}_\mu(t,E)$ in the form of movies online
\cite{SM}. In contrast to the 1D model, the masses of the atoms $m_i$ are
different in the hetero-atomic molecular junctions. This leads to the fact
that the eigenchannel displacements $\tilde{\boldsymbol{Q}}_\mu(E)$ are no
longer proportional to the eigenchannels
$\tilde{\boldsymbol{\Psi}}_\mu(E)$. Below, we will adjust the real-valued
scaling factor $s(E)$ of Eq.~(\ref{eq:QE}) for an optimized visualization of
eigenchannel displacements at each energy $E$. In this way, the vectors
$\mathrm{Re}\tilde{\boldsymbol{Q}}_\mu(E)$ at different energies should only
be compared on qualitative grounds, while they are fully comparable at a
certain fixed energy. Since it should be obvious in which situation we mean
the genuine eigenchannels $\tilde{\boldsymbol{\Psi}}_\mu(E)$ as compared to
the eigenchannel displacements $\tilde{\boldsymbol{Q}}_\mu(E)$, we do not
clearly distinguish them anymore and often simply refer to both as
``eigenchannels'' in the following. For convenience we will henceforth
furthermore omit all energy arguments.

\subsubsection{Gold dimer contact}

\begin{figure}[t]
\includegraphics[width=1.0\columnwidth]{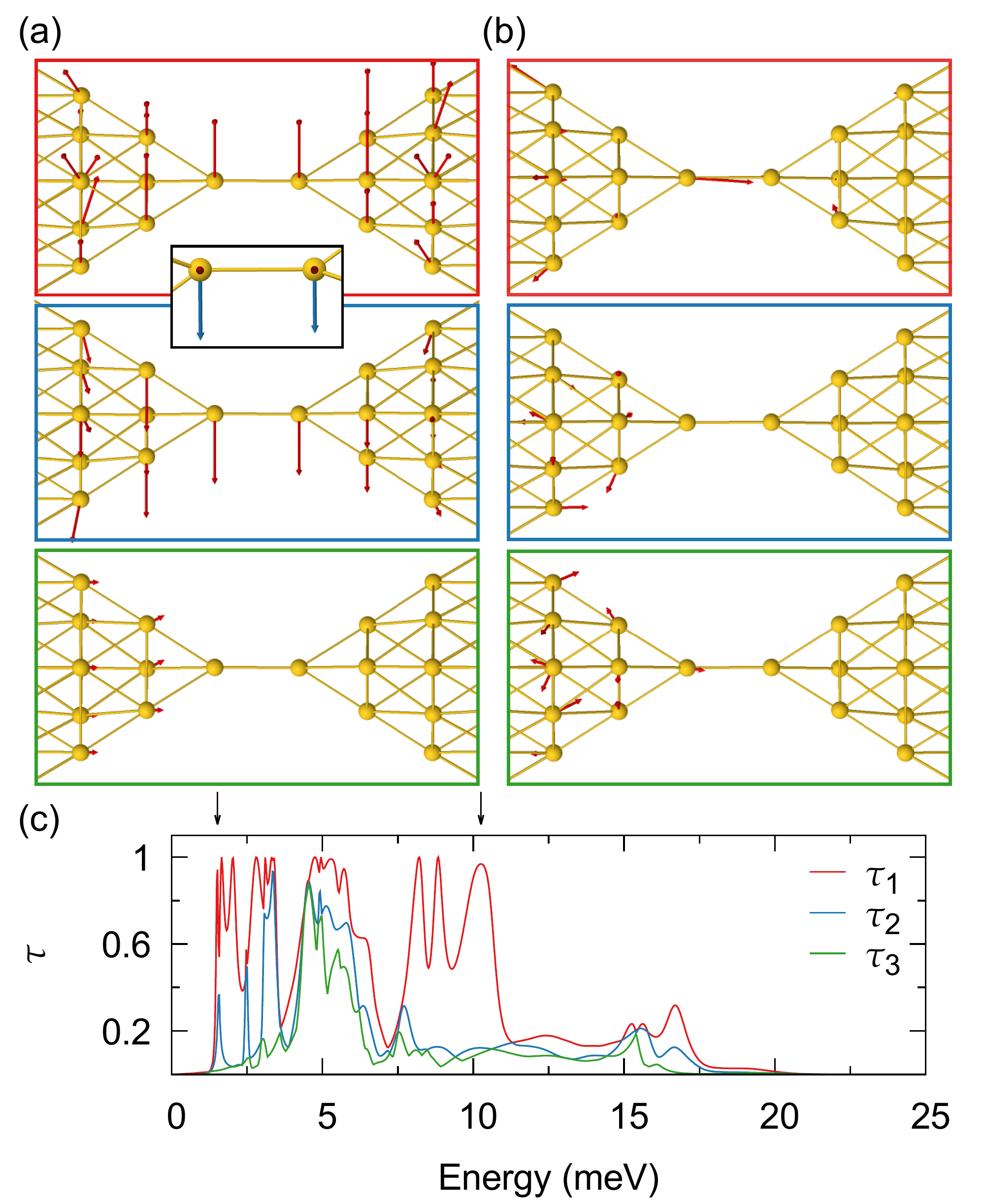}
\caption{(a) Transmission eigenchannels for a gold dimer contact at an energy
  of $1.5$ meV. We show the three eigenchannels with the highest
  transmissions, which are equal to $\tau_1=0.943$, $\tau_2=0.238$ and
  $\tau_3=0.002$. To better visualize the perpendicular polarizations of
  eigenchannels 1 and 2, we have inserted an inset between these two
  channels. It shows both $\mathrm{Re}\tilde{\boldsymbol{Q}}_1(E)$ and
  $\mathrm{Re}\tilde{\boldsymbol{Q}}_2(E)$ on the dimer atoms. Red are the
  displacements for eigenchannel 1, blue for eigenchannel 2, and the geometry
  has been rotated such that the displacement vectors of channel 1 point out
  of plane. (b) The same as in panel (a), but for an energy of $10.5$ meV. The
  channel transmissions are $\tau_1=0.968$, $\tau_2=0.122$ and
  $\tau_3=0.098$. (c) The three highest transmission coefficients as a
  function of energy for the gold atomic contact shown in panels (a) and
  (b). The energies of the eigenchannels considered in these two panels are
  indicated by arrows, and colored frames around the channel representations
  serve to identify the corresponding transmission values in panel (c).}
\label{fig-gold}
\end{figure}

The heat conductance of gold atomic contacts has been measured recently
\cite{Cui2017,Mosso2017}, and we have performed a detailed theoretical
analysis of the thermal transport due to both electrons and phonons in these
systems \cite{Cui2017,Kloeckner2017a}.  We focus here on a gold contact that
is one-atom thick and features a dimer in the narrowest part. The geometry,
which is shown in Fig.~\ref{fig-gold}, describes junctions with an electrical
conductance of the order of the electrical conductance quantum
$G_0=2e^2/h$. In Fig.~\ref{fig-gold}(a)--(c) we display the energy-dependent
phonon transmission together with eigenchannel representations for two
different energies, as indicated by arrows in the transmission plot.

In Fig.~\ref{fig-gold}(a) we show the three eigenchannels with the highest
transmission for an energy of $1.5$~meV. As one can see, the first two
channels correspond to modes with mainly transverse character with respect to
the transport direction, whose polarizations are rather perpendicular to each
other. For the first channel, with a nearly perfect transmission
$\tau_1\approx 1$, the atomic displacements are almost symmetric on both sides
of the junction. In contrast, for the second channel, with a transmission
$\tau_2<1$, a reduced amplitude is seen on the right part as compared to the
left one. This illustrates that the wave coming in from the left is mostly reflected at the
central part of the junction. The third channel shows a clearly longitudinal
character, and the amplitudes of atomic motion decay even more rapidly from
left to right because of the low transmission probability $\tau_3$. Due to the
small energy ($E=1.5$~meV) chosen in this example, the wavelength of atomic
motion spans the central part of the junction.

To explore the behavior of the eigenchannels at shorter phonon wavelengths, we
show in Fig.~\ref{fig-gold}(b) the three most transmissive eigenchannels for
an energy of $10.5$~meV. We find that the first mode is of a pronounced
longitudinal character at the dimer in the center of the junction, and we see
that the dimer atoms often move with opposite velocities, i.e., the
out-of-phase character is strongly enhanced as compared to the previous
in-phase motion at $E=1.5$~meV. Let us mention that, as discussed for the 1D
model, due to the smaller wavelength of the vibrational modes at $E=10.5$~meV,
the displacements at $t=0$ are not maximal for all of the atoms, but still one
can get an impression of the nature of the mode.  For the other two channels
at the energy of $E=10.5$~meV the amplitudes on the right junction side are,
as expected, substantially reduced due to the smaller transmission values
$\tau_2$ and $\tau_3$. While the third eigenchannel exhibits predominantly a
longitudinal character on the dimer atoms, no clear type can be assigned to
the second eigenchannel.

\begin{figure}[t]
\includegraphics[width=1.0\columnwidth]{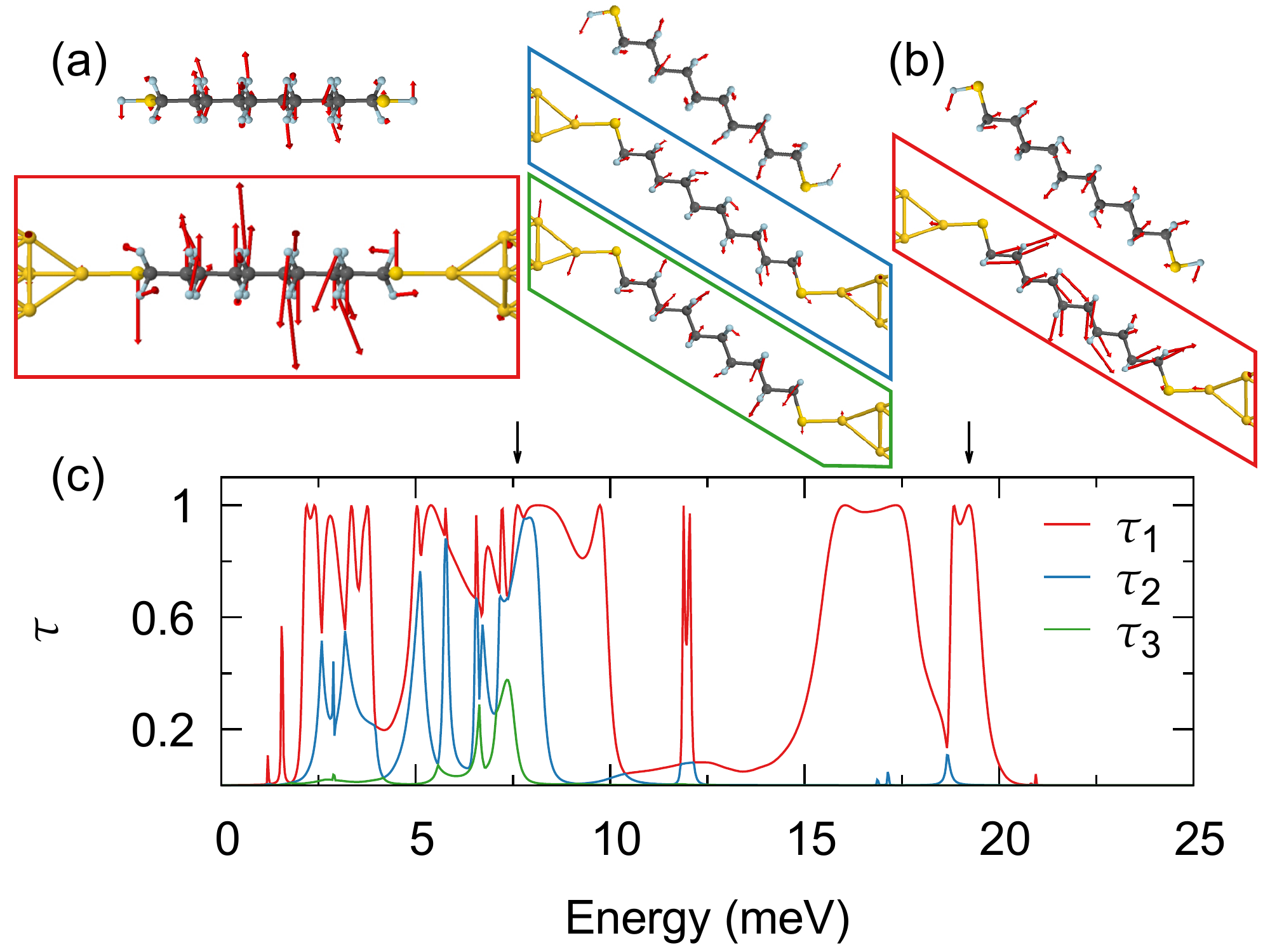}
\caption{(a) Transmission eigenchannels for an Au-decane-dithiol-Au junction
  at an energy of $7.61$~meV. We show the three eigenchannels with the highest
  transmissions of $\tau_1=1.00$, $\tau_2=0.84$ and $\tau_3=0.13$ as well as
  two vibrational modes of the free molecule at energies of $6.4$~meV and
  $9.8$~meV. (b) The most transmissive eigenchannel at an energy of
  $19.23$~meV and a vibrational mode of the free molecule at energy
  $18.76$~meV. (c) Three highest transmission eigenvalues as a function of
  energy for the Au-decane-dithiol-Au junction. The energies of the
  eigenchannels shown in panels (a) and (b) are indicated with arrows. In
  these two panels the angle of view differs for modes with out-of-plane as
  compared to in-plane character.}
\label{fig-alkane}
\end{figure}

\subsubsection{Alkane contact}

The phonon transport in molecular junctions based on alkane chains has been
studied by several theoretical groups employing different methods
\cite{Segal2003,Luo2010,Duda2011,Sadeghi2015}, and it has also been explored
experimentally in the context of many-molecule junctions
\cite{Wang2006b,Wang2007a,Losego2012,Meier2014,Majumdar2015}. In 
Ref.~\cite{Kloeckner2016} we have studied, in particular, the
length dependence of the phononic thermal conductance in single-molecule
junctions based on alkane chains. Here we focus on the analysis of a
single-molecule junction containing a dithiolated decane (i.e., an alkane
chain with 10 CH$_2$ segments) coupled to gold leads.  In
Fig.~\ref{fig-alkane}(a)--(c) we display the energy-dependent transmission
together with eigenchannel representations at the two different energies,
indicated by arrows in the transmission plot. The eigenchannels are
furthermore compared to normal modes of the isolated molecule.

In Fig.~\ref{fig-alkane}(a) we show the eigenchannels with the three highest
transmission coefficients at the energy $E=7.61$~meV. In addition, we also
show two vibrational modes of the free molecule with energies of $6.4$~meV and
$9.8$~meV. Compared to the axis through the two terminal sulfur atoms, these
modes can be described as predominantly transversal, but they can also be
classified as out-of-plane and in-plane modes, respectively, if we consider
the plane spanned by the molecular backbone of sulfur and carbon atoms. Based
on the $t=0$ snapshot, the first eigenchannel shows some similarities with the
first mode of the free molecule at $6.4$~meV, both modes having a clear
transversal, out-of-plane character. The third eigenchannel can be related to
the second mode of the free molecule at $9.8$~meV, which exhibits again a
transversal in-plane character. The relation of the second transmission
eigenchannel of in-plane type to a mode of the free molecule is, however, not
so obvious. Let us mention that we change the perspective for all modes and
eigenchannels with in-plane character as compared to those with out-of-plane
type to better visualize the atomic motions involved.

Further insight into the nature of the eigenchannels can be gained by looking
at the full time-dependent solutions \cite{SM}. In the movies available in
Ref.~\cite{SM} one can see that the first eigenchannel exhibits the
transversal, out-of-plane character that is already apparent in the static
representation of Fig.~\ref{fig-alkane}(a). For the second and third
eigenchannel, the movies reveal similar in-plane atomic motions inside the
molecule, but when compared to the sulfur-sulfur axis or those between the
two Au tip atoms, they also reveal a partially longitudinal character of the
second eigenchannel.

This example shows that an unambiguous identification of the eigenchannels
with the modes of the free molecule is not always possible. This is also
evident when considering the phase factors related to traveling waves [see the
terms $u_n(t,E)$ discussed in subsection \ref{sec:1D-chain}], which do not
appear in an isolated molecule. Nevertheless, a qualitative relation of the
eigenchannels to free modes can sometimes still be seen.

In Fig.~\ref{fig-alkane}(b) we display the most transmissive eigenchannel at
an energy of $19.23$~meV along with a vibrational mode of the free molecule
with energy $18.76$~meV that exhibits a very similar character. The
eigenchannel in this case is mainly of longitudinal, in-plane type, like the
free-molecule mode. Note the shorter wavelength of the propagating wave at the
higher energy in Fig.~\ref{fig-alkane}(b) as compared to
Fig.~\ref{fig-alkane}(a) in the static representation of the eigenchannel.

\begin{figure}[t]
\includegraphics[width=1.0\columnwidth]{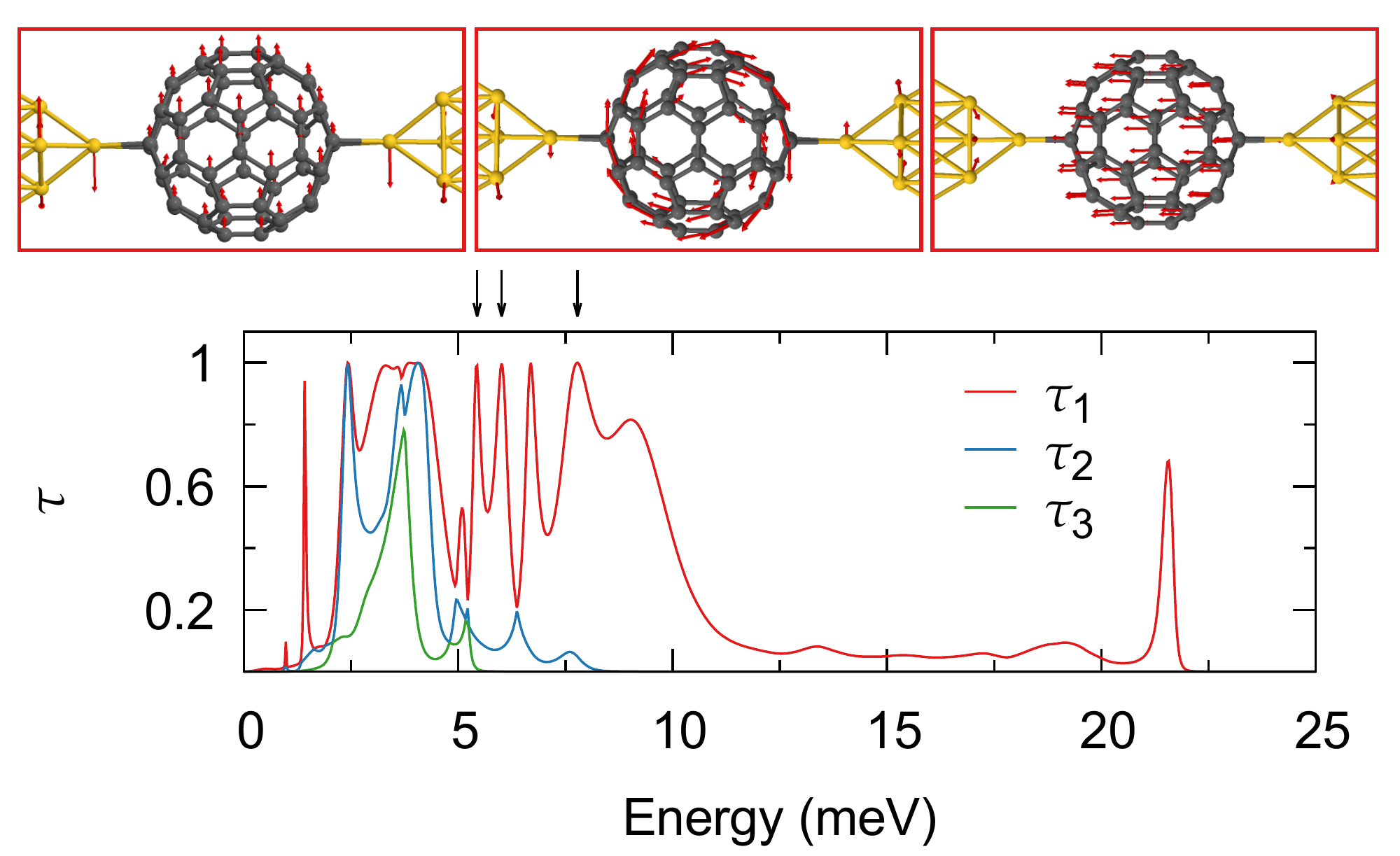}
\caption{The three largest transmission coefficients for a Au-C$_{60}$-Au
  junction as a function of energy. The most transmissive eigenchannels at
  energies of $5.4$~meV, $6.0$~meV and $7.8$~meV, as indicated by arrows in
  the transmission plot, are shown in the upper part of the figure. }
\label{fig-C60}
\end{figure}

\subsubsection{C$_{60}$ contact}

Let us now discuss the case of an Au-C$_{60}$-Au junction, see
Fig.~\ref{fig-C60}, which we have analyzed in Ref.~\cite{Kloeckner2017} in the
context of the thermoelectric figure of merit of fullerene-based
junctions. This is a very interesting case because all the vibrational modes
of the free molecule have an energy that is higher than the Au Debye
energy. So one may wonder, how phonon transport can occur in this junction and
why it is not completely suppressed. This can be nicely answered with the help
of the transmission eigenchannels.

For this purpose, we present in Fig.~\ref{fig-C60} the most transmissive
eigenchannels at energies of $5.4$~meV, $6.0$~meV, and $7.8$~meV above a plot
of the energy-dependent transmission. As one can see, all the eigenchannels
correspond to a hybridization of the vibrations of the gold atoms with the
center-of-mass motion of the C$_{60}$ molecule. The eigenchannel at $5.4$~meV
possesses a transversal character, where the molecule moves up and down as a
whole, and it also involves transversal motions of the gold atoms in the
electrode tips. The eigenchannel at $6.0$~meV involves a rotation of the
molecule that is again coupled to transversal vibrations of the gold
atoms. Finally, the eigenchannel at $7.8$~meV involves a longitudinal
center-of-mass motion of the C$_{60}$ that is coupled to a predominantly
longitudinal movement of the Au tip atoms. Let us point out that in the movies
of the eigenchannels, a small deformation of the C$_{60}$ molecule can also be
seen in all three examples in addition to the main center-of-mass motion
highlighted by the static pictures \cite{SM}.

\begin{figure}[t]
\includegraphics[width=1.0\columnwidth]{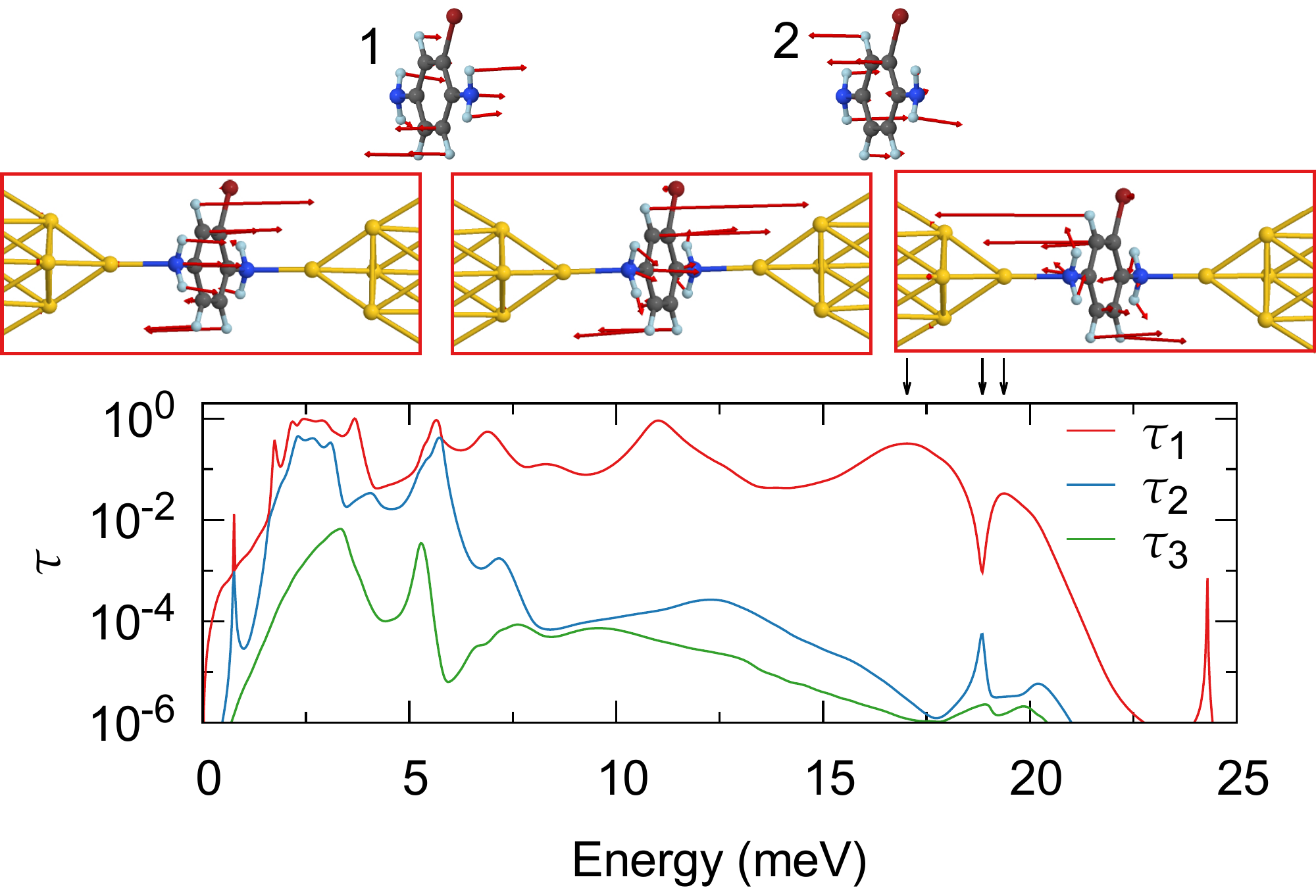}
\caption{The three highest eigenchannel transmissions as a function of energy
  for a Au-2-bromo-1,4-diaminobenzene-Au junction. Above the graph, one can
  see the eigenchannels with the highest transmission for the three different
  energies of $17.03$~meV, $18.85$~meV, and $19.37$~meV. The energies are
  selected by peaks of $\tau_1$ in the transmission plot and are indicated by
  corresponding arrows. Above the
  eigenchannels we show the vibrational modes of the free molecule that are
  responsible for the destructive interference. }
\label{fig-Fano}
\end{figure}

\subsubsection{Brominated benzene-diamine contact}

Another important example, in which the eigenchannel concept provides a
better understanding, is the case when destructive interference effects
occur in the phonon transport. We have shown in Ref.~\cite{Kloeckner2017b}
that the introduction of substituents in benzene molecules can lead to the
appearance of destructive interferences in corresponding single-molecule
junctions based on Au electrodes, despite their low Debye energy. The
interference effects are reflected in the appearance of antiresonances in the
phonon transmission.

In Fig.~\ref{fig-Fano} we show the energy dependence of the three largest
eigenchannel transmissions of an Au-2-bromo-1,4-diaminobenzene-Au junction
studied in Ref.~\cite{Kloeckner2017b}, in which an antiresonance appears at
around $19$~meV. We have attributed this antiresonance to the interference of
the two out-of-plane modes of the free molecule that lie close in energy at
$15.61$~meV and $20.07$~meV and that are shown in the upper part of
Fig.~\ref{fig-Fano}. In that figure we also display the most transmissive
eigenchannel for three different energies, which dominates the phonon
transport. As can be seen, this eigenchannel exhibits indeed a character that
closely resembles those of the two vibrational modes of the free molecule at
similar energies. Additionally, in the static picture of the eigenchannel
shown in Fig.~\ref{fig-Fano}, one can see a jump of $\pi$ in the phase of
molecular motion as compared to the Au reference atoms, which is reflected by
the fact that the arrows on the molecule point in opposite directions at
energies above and below the antiresonance. This jump in the phase is a
well-known phenomenon that accompanies destructive interference in the
electronic transport \cite{Solomon2010} and, in general, in the Fano model
\cite{Joe2006}.

Our example illustrates how the eigenchannel concept helps to identify the
molecular origin of destructive quantum interference. This is particularly
useful in cases in which is not easy to figure out the vibrational modes
responsible for the interference phenomenon, because of the presence of many
other modes in that energy region or because the energies of the vibrations of
the free molecule are strongly renormalized by the hybridization with the
metallic leads when the molecule is connected to the electrodes.

\section{Conclusions} \label{sec-Concl}

In this work, and in analogy with what is done in electronic transport, we
have presented a method to obtain the transmission eigenchannels from
NEGF-based calculations of the coherent phonon transport. In particular, we
have shown that this method can be combined with ab initio simulations to
provide an insight into phonon transport that cannot simply be obtained from
the analysis of the transmission probabilities. We have illustrated this
approach with the analysis of the phonon eigenchannels in realistic
atomic-scale junctions, including single-atom and single-molecule
junctions. Moreover, we have discussed different ways to visualize these
eigenchannels by means of static and time-dependent representations. We
believe that the procedure presented in this work will become a valuable tool
for the analysis of coherent phonon transport in a great variety of nanoscale
systems and devices.

\section{Acknowledgments}

J.C.K.\ and F.P.\ thank M. B\"urkle for stimulating discussions in the initial
phase of this work. In addition, both gratefully acknowledge funding from the
Carl Zeiss foundation, the Junior Professorship Program of the Ministry of
Science, Research and the Arts of the state of Baden W\"urttemberg, and the
Collaborative Research Center (SFB) 767 of the German Research Foundation
(DFG). J.C.C.\ thanks the Spanish Ministry of Economy and Competitiveness
(Contract No.\ FIS2017-84057-P) for financial support as well as the DFG and
SFB 767 for sponsoring his stay at the University of Konstanz as Mercator
Fellow. An important part of the numerical modeling was carried out on the
computational resources of the bwHPC program, namely the bwUniCluster and the
JUSTUS HPC facility.

\end{document}